\documentclass[twoside]{article}
\usepackage{fleqn,espcrc2}

\usepackage{graphicx}
\usepackage{epsf}
\usepackage[figuresright]{rotating}
\newcommand{\AmS}{{\protect\the\textfont2
  A\kern-.1667em\lower.5ex\hbox{M}\kern-.125emS}}
\def\Frac#1#2{\frac{\displaystyle{#1}}{\displaystyle{#2}}}
\def\lsim{\raise0.3ex\hbox{$\;<$\kern-0.75em\raise-1.1ex\hbox{$\sim\;$}}}
\def\gsim{\raise0.3ex\hbox{$\;>$\kern-0.75em\raise-1.1ex\hbox{$\sim\;$}}}
\def\npb#1#2#3{    {\it Nucl. Phys. }{\bf B #1} (19#2) #3}
\def\npbII#1#2#3{    {\it Nucl. Phys. }{\bf B #1} (20#2) #3}
\def\plb#1#2#3{    {\it Phys. Lett. }{\bf B #1} (19#2) #3}
\def\plbII#1#2#3{    {\it Phys. Lett. }{\bf B #1} (20#2) #3}
\def\prd#1#2#3{    {\it Phys. Rev. }{\bf D #1} (19#2) #3}
\def\prdII#1#2#3{    {\it Phys. Rev. }{\bf D #1} (20#2) #3}
\def\prep#1#2#3{   {\it Phys. Rep. }{\bf #1} (19#2) #3}
\def\prl#1#2#3{    {\it Phys. Rev. Lett. }{\bf #1} (19#2) #3}

\def\zpc#1#2#3{    {\it Zeit. f\"ur Physik }{\bf C #1} (19#2) #3}
\def\ibid#1#2#3{   {\it ibid. }{\bf #1} (19#2) #3}
\newcommand\jhep[3]  {{\it J. High Energy Phys.\ }{\bf #1} (#2) #3}

\title{CP Violation in SUSY}

\author{A. Masiero and O. Vives\thanks{Present address Dep F\'{\i}sica Te\`orica, U de Val\`encia, Spain} \address{SISSA, Via Beirut 2--4, 
34013 Trieste, Italy and\\ 
INFN, sez. di Trieste, Trieste, Italy}%
}
       
\begin{document}

\begin{abstract}
Supersymmetry exhibts new sources of CP violation. We discuss the
implications of these new contributions to CP violation both in the K and
B physics. We show that CP violation puts severe constraints on low energy
SUSY, but it represents also a promising ground to look for signals of new
physics.
\vspace{1pc}
\end{abstract}

\maketitle
\section{CP VIOLATION IN SUSY}

CP violation has major potentialities to exhibit manifestations of new physics
beyond the standard model.
Indeed, it is quite a general feature that new physics possesses
new CP violating phases in addition to the
Cabibbo--Kobayashi--Maskawa (CKM) phase $\left(\delta_{\rm CKM}\right)$
or, even in those cases where this does not occur, $\delta_{\rm CKM}$
shows up in interactions of the new particles, hence with potential departures
from the SM expectations. Moreover, although the SM is able to account for the
observed CP violation in the kaon system, we cannot say that we have tested so
far the SM predictions for CP violation. The detection of CP violation in $B$
physics will constitute a crucial test of the standard CKM picture within the
SM. Again, on general grounds, we expect new physics to provide departures from
the SM CKM scenario for CP violation in $B$ physics. A final remark on reasons
that make us optimistic in having new physics playing a major role in CP
violation concerns the matter--antimatter asymmetry in the universe. Starting
from a baryon--antibaryon symmetric universe, the SM is unable to account for
the observed baryon asymmetry. The presence of new CP--violating contributions
when one goes beyond the SM looks crucial to produce an efficient mechanism for
the generation of a satisfactory $\Delta$B asymmetry.

The above considerations apply well to the new physics represented by
low--energy supersymmetric extensions of the SM. Indeed, as we will see below,
supersymmetry introduces CP violating phases in addition to
$\delta_{\rm CKM}$ and, even if one envisages particular situations
where such extra--phases vanish, the phase $\delta_{\rm CKM}$ itself
leads to new CP--violating contributions in processes where SUSY particles are
exchanged. CP violation in $B$ decays has all potentialities to exhibit
departures from the SM CKM picture in low--energy SUSY extensions, although, as
we will discuss, the detectability of such deviations strongly depends on the
regions of the SUSY parameter space under consideration.

In any MSSM, at least two new ``genuine'' SUSY CP--violating phases are 
present. They
originate from the SUSY parameters $\mu$, $M$, $A$ and $B$. The first of these
parameters is the dimensionful coefficient of the $H_u H_d$ term of the
superpotential. The remaining three parameters are present in the sector that
softly breaks the N=1 global SUSY. $M$ denotes the common value of the gaugino
masses, $A$ is the trilinear scalar coupling, while $B$ denotes the bilinear
scalar coupling. In our notation, all these three parameters are
dimensionful. The simplest way to see which combinations of the phases of these
four parameters are physical \cite{Dugan} is to notice that for vanishing
values of $\mu$,  $M$, $A$ and $B$ the theory possesses two additional
symmetries \cite{Dimopoulos}. Indeed, letting $B$ and $\mu$ vanish, a $U(1)$
Peccei--Quinn symmetry originates, which in particular rotates $H_u$ and $H_d$.
If $M$, $A$ and $B$ are set to zero, the Lagrangian acquires a continuous
$U(1)$ $R$ symmetry. Then we can consider  $\mu$,  $M$, $A$ and $B$ as spurions
which break the $U(1)_{PQ}$ and $U(1)_R$ symmetries. In this way, the question
concerning the number and nature of the meaningful phases translates into the
problem of finding the independent combinations of the four parameters which
are invariant under $U(1)_{PQ}$ and $U(1)_R$ and determining their independent
phases. There are three such independent combinations, but only two of their
phases are independent. We use here the commonly adopted choice:
\begin{equation}
  \label{CMSSMphases}
  \varphi_A = {\rm arg}\left( A^* M\right), \qquad
  \varphi_B = {\rm arg}\left( B^* M\right).
\end{equation}
where also ${\rm arg}\left( B \mu\right) = 0$, i.e. 
$\varphi_\mu= - \varphi_B$.

The main constraints on $\varphi_A$ and $\varphi_B$ come from their 
contribution to
the electric dipole moments of the neutron and of the electron. For instance,
the effect of $\varphi_A$ and $\varphi_B$ on the electric and chromoelectric dipole
moments of the light quarks ($u$, $d$, $s$) lead to a contribution to
$d^e_N$ of 
order \cite{EDMN}
\begin{equation}
  \label{EDMNMSSM}
  d^e_N \sim 2 \left( \Frac{100 {\rm GeV}}{\tilde{m}}\right)^2 \sin \varphi_{A,B}
  \times 10^{-23} {\rm e\, cm},
\end{equation}
where $\tilde{m}$ here denotes a common mass for squarks and gluinos. The
present experimental bound, $d^e_N < 1.1 \times 10^{-25}$ e cm, implies that
$\varphi_{A,B}$ should be $<10^{-2}$, unless one pushes SUSY masses up to 
${\cal{O}}$(1 TeV). A possible caveat to such an argument calling for a 
fine--tuning of
$\varphi_{A,B}$ is that uncertainties in the estimate of the hadronic matrix
elements could relax the severe bound in Eq.~(\ref{EDMNMSSM}) \cite{Ellis}.

In view of the previous considerations, most authors dealing with the MSSM
prefer to simply put $\varphi_A$ and $\varphi_B$ equal to zero. Actually, one may
argue in favor of this choice by considering the soft breaking sector of the
MSSM as resulting from SUSY breaking mechanisms which force $\varphi_A$ and
$\varphi_B$ to vanish. For instance, it is conceivable that both $A$ and $M$
originate from one same source of $U(1)_R$ breaking. Since $\varphi_A$ ``measures''
the relative phase of $A$ and $M$, in this case it would ``naturally''vanish. 
In some specific models, it has been shown \cite{Dine} that through an 
analogous mechanism also $\varphi_B$ may vanish.

If $\varphi_A=\varphi_B=0$, then the novelty of SUSY in CP violating 
contributions
merely arises from the presence of the CKM phase in loops where SUSY particles
run \cite{CPSUSY}. The crucial point is that the usual GIM suppression, which
plays a major role in evaluating $\varepsilon_K$ and 
$\varepsilon^\prime/\varepsilon$ in the SM, in the MSSM case (or more exactly 
in the CMSSM) is replaced by a super--GIM cancellation which has the same
``power'' of suppression as the original GIM (see previous section). Again,
also in
the CMSSM, as it is the case in the SM, the smallness of $\varepsilon_K$ and
$\varepsilon^\prime/\varepsilon$ is guaranteed not by the smallness of
$\delta_{\rm CKM}$, but
rather by the small CKM angles and/or small Yukawa couplings. By the same
token, we do not expect any significant departure of the CMSSM from the SM
predictions also concerning CP violation in $B$ physics. As a matter of fact,
given the large lower bounds on squark and gluino masses, one expects
relatively tiny contributions of the SUSY loops in $\varepsilon_K$ or
$\varepsilon^\prime/\varepsilon$ in comparison with the normal $W$ loops of 
the SM. Let us be more detailed on this point.

In the CMSSM, the gluino exchange contribution
to FCNC is subleading with respect to chargino ($\chi^\pm$) and charged
Higgs ($H^\pm$) exchanges. Hence, when dealing with CP violating FCNC
processes in the CMSSM with $\varphi_A=\varphi_B=0$, one can confine the analysis
 to $\chi^\pm$
and $H^\pm$ loops. If one takes all squarks to be degenerate in mass and
heavier than $\sim 200$ GeV, then $\chi^\pm-\tilde q$ loops are obviously
severely penalized with respect to the SM $W^+$--$q$ loops (remember that at the
vertices the same CKM angles occur in both cases).

The only chance for the CMSSM to produce some sizeable departure from the SM
situation in CP violation is in the particular region of the parameter space
where one has light $\tilde q$, $\chi^\pm$ and/or $H^\pm$. The best
candidate (indeed the only one unless 
$\tan \beta \sim m_t/m_b$) for a light squark is the stop. Hence one can
ask the following question: can the CMSSM present some novelties in CP--violating
phenomena when we consider $\chi^+$--$\tilde t$ loops with light $\tilde t$,
$\chi^+$ and/or $H^+$?

Several analyses in the literature tackle the above question or, to be more
precise, the more general problem of the effect of light $\tilde t$
and $\chi^+$ 
on FCNC processes \cite{refbrignole,mpr,branco}. A first important
observation concerns the
relative sign of the $W^+$--$t$ loop with respect to the  $\chi^+$--$\tilde t$ 
and $H^+$--$t$ contributions. As it is well known, the latter contribution 
always
interferes positively with the SM one. Interestingly enough, in the region of
the MSSM parameter space that we consider here, also the $\chi^+$--$\tilde t$
contribution interferes constructively with the SM contribution. The second
point regards the composition of the lightest chargino, i.e. whether the
gaugino or higgsino component prevails. This is crucial since the light stop is
predominantly $\tilde t_R$ and, hence, if the lightest chargino is mainly a
wino, it couples to $\tilde t_R$ mostly through the $LR$ mixing in the stop
sector. Consequently, a suppression in the contribution to box diagrams going
as $\sin^4 \theta_{LR}$ is present ($\theta_{LR}$ denotes the mixing angle
between
the lighter and heavier stops). On the other hand, if the lightest chargino is
predominantly a higgsino (i.e. $M_2 \gg \mu$ in the chargino mass matrix), then
the $\chi^+$--lighter $\tilde t$ contribution grows. In this case, 
contributions
$\propto \theta_{LR}$ become negligible and, moreover, it can be shown that
they are independent on the sign of $\mu$. A detailed study is provided in
reference \cite{mpr,branco}. For instance, for $M_2/\mu=10$, they find that 
the inclusion of
the SUSY contribution to the box diagrams doubles the usual SM contribution for
values of the lighter $\tilde t$ mass up to $100$--$120$ GeV, using $\tan \beta
=1.8$, $M_{H^+}=100$ TeV, $m_\chi=90$ GeV and the mass of the heavier $\tilde
t$ of 250 GeV. However, if $m_\chi$ is pushed up to 300 GeV, the  
$\chi^+$--$\tilde t$ loop yields a contribution which is roughly 3 times less 
than in the
case $m_\chi=90$ GeV, hence leading to negligible departures from the SM
expectation. In the cases where the SUSY contributions are sizeable, one
obtains relevant restrictions on the $\rho$ and $\eta$ parameters of the CKM
matrix by making a fit of the parameters $A$, $\rho$ and $\eta$ of the CKM
matrix and of the total loop contribution to the experimental values of
$\varepsilon_K$ and $\Delta M_{B_d}$. For instance, in the above--mentioned
case in which the SUSY loop contribution equals the SM $W^+$--$t$ loop, hence giving
a total loop contribution which is twice as large as in the pure SM case,
combining the $\varepsilon_K$ and $\Delta M_{B_d}$ constraints leads to a
region in the $\rho$--$\eta$ plane with $0.15<\rho<0.40$ and $0.18<\eta<0.32$,
excluding negative values of $\rho$.

In conclusion, the situation concerning CP violation in the MSSM case with
$\varphi_A=\varphi_B=0$ and exact universality in the soft--breaking sector can be
summarized in the following way: the MSSM does not lead to any significant
deviation from the SM expectation for CP--violating phenomena as $d_N^e$,
$\varepsilon_K$, $\varepsilon^\prime/\varepsilon$ and CP violation in $B$ physics; the only
exception to this statement concerns a small portion of the MSSM
parameter space 
where a very light $\tilde t$ ($m_{\tilde t} < 100$ GeV) and $\chi^+$
($m_\chi \sim 90$ GeV) are present. In this latter particular situation,
sizeable SUSY contributions to $\varepsilon_K$ are possible and, consequently,
major restrictions in the $\rho$--$\eta$ plane can be inferred. Obviously, CP
violation in $B$ physics becomes a crucial test for this MSSM case with very
light $\tilde t$ and $\chi^+$. Interestingly enough, such low values of SUSY
masses are at the border of the detectability region at LEP II.

In next Section, we will move to the case where, still keeping the
minimality of the model, we switch on the new CP violating phases.
Later on we will give up also the strict minimality related to the absence
of new flavor structure in the SUSY breaking sector and we will see that,
in those more general contexts, we can expect SUSY to significantly depart
from the SM predictions in  CP violating phenomena. 

\section{FLAVOR BLIND SUSY BREAKING AND CP VIOLATION}
\label{sec:flavor-blind}

We have seen in the previous section that in any MSSM there are additional 
phases which can cause deviations from the predictions of the SM in CP 
violation experiments. In fact, in the CMSSM, there are already two new phases 
present, Eq.(\ref{CMSSMphases}), and for most of the MSSM parameter 
space, 
the experimental bounds on the electric dipole moments (EDM) of the electron 
and neutron constrain these phases to be at most ${\cal{O}}(10^{-2})$.  
However, in the last few years, the possibility of having non--zero SUSY phases
has again attracted a great deal of attention. Several new mechanisms have 
been proposed to suppress supersymmetric contributions to EDMs below the 
experimental bounds while allowing SUSY phases ${\cal{O}}(1)$. 
Methods of suppressing the EDMs 
consist of cancellation of various SUSY contributions among themselves 
\cite{cancel}, non universality of the soft breaking parameters at the 
unification scale \cite{non-u} and approximately degenerate heavy sfermions 
for the first two generations \cite{heavy}. 
In the presence of one of these mechanisms, large supersymmetric phases are
naturally expected and EDMs should be generally close to the experimental 
bounds. \footnote{In a more general (and maybe more natural) MSSM
there are many other CP violating phases \cite{124} that contribute to CP 
violating observables.}

In this section we will study the effects of these phases in CP violation
observables as $\varepsilon_K$, $\varepsilon^\prime/\varepsilon$ and $B^0$ 
CP asymmetries. Following our work of ref. \cite{flavor} it is clear that 
the presence of large SUSY 
phases is not enough to produce sizeable supersymmetric contributions to 
these observables. In fact, {\it in the absence of the CKM phase, a general 
MSSM with all possible phases in the soft--breaking terms, but no new flavor 
structure beyond the usual Yukawa matrices, can never give a sizeable 
contribution to $\varepsilon_K$, $\varepsilon^\prime/\varepsilon$ or hadronic 
$B^0$ CP asymmetries}. However, we will see in the next section, that  
as soon as one introduces some new flavor structure in the soft SUSY--breaking 
sector, even if the CP violating phases are flavor independent, it is indeed 
possible to get sizeable CP contribution for large SUSY phases and 
$\delta_{CKM}=0$.
Then, we can rephrase our sentence above in a different way: {\it A new result 
in hadronic $B^0$ CP asymmetries in the framework of supersymmetry would be 
a direct proof of the existence of a completely new flavor structure in the 
soft--breaking terms}. This means that $B$--factories will probe the flavor 
structure of the supersymmetry soft--breaking terms even before the direct 
discovery of the supersymmetric partners \cite{flavor}. 

\section{CP VIOLATION IN THE PRESENCE OF NEW FLAVOR STRUCTURES}
\label{sec:newflavor}
In section \ref{sec:flavor-blind}, we have shown that CP violation effects are 
always small in models with flavor blind soft--breaking terms. 
However, as soon as one introduces some new flavor structure in the soft 
breaking sector, it is indeed possible to get sizeable CP contribution 
for large SUSY phases and $\delta_{CKM}=0$ \cite{non-u,brhlik2,newflavor}.
To show this, we will mainly concentrate in new supersymmetric 
contributions to $\varepsilon^\prime/\varepsilon$.
 
In the CMSSM, the SUSY contribution to $\varepsilon^\prime/\varepsilon$ is
small \cite{giudice,flavor}. However in a MSSM with a more general framework of
flavor structure it is relatively easy to obtain larger SUSY effects to
$\varepsilon^\prime/\varepsilon$. In ref. \cite{murayama} it was shown
that such large SUSY contributions arise once one assumes that:  i)
hierarchical quark Yukawa matrices are protected by flavor symmetry, ii) a
generic dependence of Yukawa matrices on Polonyi/moduli fields is present
(as expected in many supergravity/superstring theories), iii) the Cabibbo
rotation originates from the down--sector and iv) the phases are of order
unity. In fact, in \cite{murayama}, it was illustrated how the observed
$\varepsilon^\prime/\varepsilon$ could be mostly or entirely due to the
SUSY contribution. 

The universality of the breaking is a strong assumption and is known not
to be true in many supergravity and string inspired models \cite{BIM}.
In these models, we expect at least some non--universality in the squark
mass matrices or tri--linear terms at the supersymmetry breaking scale.
Hence, sizeable flavor--off-diagonal entries will appear in the squark
mass matrices.
In this regard, gluino contributions to $\varepsilon^\prime/\varepsilon$
are especially sensitive to $(\delta^{d}_{12})_{LR}$; even 
$|{\rm Im}(\delta^{d}_{12})_{LR}^{2}| \sim 10^{-5}$ gives a significant 
contribution to $\varepsilon^\prime/\varepsilon$ while keeping the
contributions from this MI to $\Delta m_{K}$ and $\varepsilon_K$ well bellow 
the phenomenological bounds. The situation is the opposite for $L$--$L$
and $R$--$R$ mass insertions; the stringent bounds on 
$(\delta^{d}_{12})_{LL}$ and $(\delta^{d}_{12})_{RR}$ 
from $\Delta m_{K}$ and $\varepsilon_K$ prevent them to contribute 
significantly to $\varepsilon^\prime/\varepsilon$. 

The LR squark mass matrix has the same flavor structure as the fermion 
Yukawa matrix and both, in fact, originate from the superpotential couplings. 
It may be appealing to invoke the presence of an underlying flavor symmetry 
restricting 
the form of the Yukawa matrices to explain their hierarchical forms. Then, 
the LR mass matrix is expected to have a very similar form as the Yukawa
matrix. Indeed, we expect the components of the LR mass matrix to be
roughly the SUSY breaking scale (e.g., the gravitino mass) times the
corresponding component of the quark mass matrix. However, there is no
reason for them to be simultaneously diagonalizable based on this general
argument.  
To make an order of magnitude estimate, we take the down quark mass matrix 
for the first and second generations to be (following our assumption iii)),
\begin{equation}
        Y^{d} v_1 \simeq \left( \begin{array}{cc}
                m_{d} & m_{s} V_{us} \\
                 & m_{s}
        \end{array} \right),
\end{equation}
where the (2,1) element is unknown due to our lack of knowledge on 
the mixings among right--handed quarks (if we neglect small terms 
$m_d V_{cd}$).  Based on the general 
considerations on the LR mass matrix above, we expect
\begin{equation}
        {m^{2}}^{(d)}_{LR} \simeq m_{3/2}  \left( \begin{array}{cc}
                a m_{d} & b m_{s} V_{us} \\
                 & c m_{s}
        \end{array} \right) ,
\end{equation}
where $a$, $b$, $c$ are constants of order unity.  Unless $a=b=c$ 
exactly, $M_{d}$ and $m^{2,d}_{LR}$ are not simultaneously 
diagonalizable and we find
\begin{eqnarray}
       & (\delta^{d}_{12})_{LR} \simeq \Frac{m_{3/2} m_{s} 
        V_{us}}{m_{\tilde{q}}^{2}} \nonumber = 2 \times 10^{-5} \cdot \\ 
       & \left( \Frac{m_{s}(M_{Pl})}{\rm 50~MeV}\right)
        \left(\Frac{m_{3/2}}{m_{\tilde{q}}}\right)
        \left(\Frac{\rm 500~GeV}{m_{\tilde{q}}}\right)
        \label{eq:estimate}
\end{eqnarray}
It turns out that, following the simplest implementation along
the lines of the above described idea, the amount of flavor changing LR
mass insertion in the s and d--squark propagator results to roughly
saturate the bound from $\varepsilon^\prime/\varepsilon$ if a SUSY phase
of order unity is present \cite{murayama}. 

This line of work has received a great deal of attention in recent times,
after the last experimental measurements of $\varepsilon^\prime/\varepsilon$
in KTeV and NA31 \cite{KTeV,NA31}. The effects of non--universal $A$ terms in
CP violation experiments were previously analyzed by Abel and Frere 
\cite{abel} and after this new measurement discussed in many different works
\cite{non-u}. In the following we show a complete realization of the above 
Masiero--Murayama (MM) mechanism from a Type I string--derived model
recently presented by one of the authors \cite{KKV}.

\subsection{TYPE I STRING MODEL AND $\varepsilon^\prime/\varepsilon$}
\label{sec:type1}

In first place we explain our starting model, which is based on type I 
string models. Our purpose is to study explicitly CP violation effects in 
models with non--universal gaugino masses and $A$--terms.
Type I models can realize such initial conditions.
These models contain nine--branes and three types of 
five--branes ($5_a$, $a=1,2,3$).
Here we assume that the gauge group $SU(3)\times U(1)_Y$ is on a 9--brane 
and the gauge group $SU(2)$ on the $5_1$--brane like in 
Ref.~\cite{brhlik2,ibrahim}, in order to get non--universal gaugino masses 
between $SU(3)$ and $SU(2)$.
We call these branes the $SU(3)$--brane and the $SU(2)$--brane, 
respectively.
\vskip 0.25cm

Chiral matter fields correspond to open strings spanning between 
branes. Thus, they must be assigned accordingly to their quantum numbers.
For example, the chiral field corresponding to the 
open string between the $SU(3)$ and $SU(2)$ branes has 
non--trivial representations under both $SU(3)$ and $SU(2)$, 
while the chiral field corresponding to the open string, 
which starts and ends on the $SU(3)$--brane, should be 
an $SU(2)$--singlet.
\vskip 0.25cm
There is only one type of the open string that spans between the 9 and 
5--branes, that we denote as the $C^{95_1}$.
However, there are three types of open strings which start and end on the 
9--brane, that is, the $C_i^9$ sectors (i=1,2,3), corresponding to the $i$--th
complex compact dimension among the three complex dimensions.
If we assign the three families to the different $C_i^9$ sectors 
we obtain non--universality in the right--handed sector.
Notice that, in this model, we can not derive non--universality for the 
squark doublets, i.e. the left--handed sector.
In particular, we assign the $C^{9}_1$ sector to the third family and
the $C^{9}_3$ and $C^{9}_2$, to the first and second families, respectively.
\vskip 0.25cm

Under the above assignment of the gauge multiplets and 
the matter fields, soft SUSY breaking terms are obtained,  
following the formulae in Ref.~\cite{typeI}.
The gaugino masses are obtained 
\begin{eqnarray}
\label{gaugino}
M_3 & = & M_1 = \sqrt 3 m_{3/2} \sin \theta  e^{-i\alpha_S}, \\
M_2 & = &  \sqrt 3 m_{3/2} \cos \theta \Theta_1 e^{-i\alpha_1}.
\end{eqnarray}
While the $A$--terms are obtained as 
\begin{equation}
A_{C_1^9}= -\sqrt 3 m_{3/2} \sin \theta e^{-i\alpha_S}=-M_3,
\label{A-C1}
\end{equation}
for the coupling including $C_1^{9}$, i.e. the third family, 
\begin{eqnarray}
A_{C_2^9}= -M_3 -\sqrt 3 m_{3/2}
\cos \theta (\Theta_1 e^{-i\alpha_1}- \Theta_2 e^{-i\alpha_2}),
\nonumber \\
\label{A-C2}
A_{C_3^9}= -M_3 -\sqrt 3 m_{3/2}
\cos \theta (\Theta_1 e^{-i\alpha_1}- \Theta_3 e^{-i\alpha_3}),\nonumber \\
\end{eqnarray}
for the second and first families.
Here $m_{3/2}$ is the gravitino mass, $\alpha_S$ and $\alpha_i$ are 
the CP phases of the F--terms of the dilaton field $S$ and 
the three moduli fields $T_i$, and $\theta$ and $\Theta_i$ are 
goldstino angles, and we have the constraint, $\sum \Theta_i^2=1$.
\vskip 0.25cm 
Thus, if quark fields correspond to different 
$C_i^9$ sectors, we have non--universal A--terms.
We obtain the following  trilinear SUSY breaking matrix, 
$(Y^A)_{ij}=(Y)_{ij}(A)_{ij}$, 
\begin{equation}
Y^A = \left(\begin{array}{ccc}
 & & \\ &Y_{ij}&  \\ & &  \end{array}
\right) \cdot 
\left(\begin{array}{ccc}
A_{C^9_3} & 0 & 0 \\ 0 & A_{C^9_2} & 0 \\ 0 & 0 & A_{C^9_1} \end{array}
\right)
\end{equation}
\vskip 0.25cm
In addition, soft scalar masses for quark doublets and 
the Higgs fields are obtained, 
\begin{equation}
\label{doublets}
m_{C^{95_1}}^2=m_{3/2}^2(1-{3 \over 2}\cos^2 \theta(1- 
\Theta_1^2)).
\end{equation}
The soft scalar masses for quark singlets are obtained as
\begin{equation}
\label{singlets}
m_{C_i^9}^2=m_{3/2}^2(1-3\cos^2 \theta \Theta^2_i),
\end{equation}
if it corresponds to  the $C_i^{9}$ sector.
\vskip 0.25cm

Now, below the string or SUSY breaking scale, this model is simply a MSSM 
with non--trivial soft--breaking terms
from the point of view of flavor. Scalar mass matrices and tri--linear terms
have completely new flavor structures, as opposed to the super--gravity 
inspired CMSSM or the SM, where the only connection 
between different generations is provided by the Yukawa matrices.

This model includes, in the quark sector, 7 different structures of flavor,
$M_{Q}^2$, $M_{U}^2$, $M_{D}^2$, $Y_d$, $Y_u$, $Y^A_d$ and $Y^A_u$. 
From these matrices, $M_{Q}^2$, the squark doublet mass matrix, is 
proportional to the identity matrix, and hence trivial, then we are left 
with 6 non--trivial flavor matrices.
Notice that we have always the freedom to diagonalize the hermitian
squark mass matrices (as we have done in the previous section, 
Eqs.(\ref{doublets},\ref{singlets})) and fix some general form for the
Yukawa and tri--linear matrices. In this case, these four matrices
are completely observable, unlike in the SM or CMSSM case.

At this point, to specify completely the model, we need not only the 
soft--breaking terms but also the complete Yukawa textures.
The only available experimental information is the Cabbibo--Kobayashi--Maskawa 
(CKM) mixing matrix and the quark masses. Here, we choose our Yukawa texture 
following two simple assumptions : i) the CKM mixing matrix originates from 
the down Yukawa couplings (as done in the MM case) and ii) our Yukawa 
matrices are hermitian \cite{RRR}.
With these two assumptions we fix completely the Yukawa matrices as
$ v_1\, Y_d = {K}^\dagger\cdot M_d\cdot K$ and $ v_2\, Y_u = M_u$, 
with $M_d$ and $M_u$ diagonal quark mass matrices, $K$ the 
Cabibbo--Kobayashi--Maskawa (CKM) mixing matrix and $v= v_1 /(\cos \beta) = 
v_2 / (\sin \beta) = \sqrt{2} M_W / g$. We take $\tan \beta= v_2/v_1 = 2$ in 
the following in all numerical examples. 
In this basis we can analyze the down tri--linear matrix at the string scale,
\begin{equation}
Y^A_d =\ K^\dagger\cdot \Frac{M_d}{v_1}\cdot K\cdot
\left(\begin{array}{ccc}
A_{C^9_3} & 0 & 0 \\ 0 & A_{C^9_2} & 0 \\ 0 & 0 & A_{C^9_1} \end{array}
\right)
\end{equation}

Hence, together with the up tri--linear matrix we have our MSSM completely
defined. The next step is simply to use the MSSM Renormalization Group 
Equations \cite{RGE,BBM} to obtain the whole spectrum and couplings at the
low scale, $M_W$. The dominant effect in the tri--linear terms renormalization
is due to the gluino mass which produces the well--known alignment among
A--terms and gaugino phases. However, this renormalization is always 
proportional to the Yukawa couplings and not to the tri--linear terms. 
This implies that, in the SCKM basis, the gluino effects 
will be diagonalized in excellent approximation, while due to the different
flavor structure of the tri--linear terms large off--diagonal elements will
remain with phases ${\cal{O}}(1)$ \cite{murayama}. To see this more 
explicitly, we can roughly approximate the RGE effects at $M_W$ as,
\begin{equation}
Y^A_d = c_{\tilde{g}}m_{\tilde{g}} Y_d+c_{A} Y_d \cdot
\left(\begin{array}{ccc}
A_{C^9_3} & 0 & 0 \\ 0 & A_{C^9_2} & 0 \\ 0 & 0 & A_{C^9_1} \end{array}
\right)
\end{equation}
with $m_{\tilde{g}}$ the gluino mass and $c_{\tilde{g}}$, $c_A$ coefficients 
order 1 (typically $c_{\tilde{g}} \simeq 5$ and $c_A\simeq 1$).

We go to the SCKM basis after diagonalizing all the Yukawa matrices
(that is, $K . Y_d . K^\dagger = M_d /v_1$).
In this basis, we obtain the tri--linear couplings as,
\begin{eqnarray}
Y^A_d = \Big(c_{A}\Frac{M_d}{v_1}\cdot K\cdot \mbox{Diag }
(A_{C^9_3},A_{C^9_2},A_{C^9_1})\cdot K^\dagger \nonumber \\
+ c_{\tilde{g}}\ m_{\tilde{g}}\Frac{M_d}{v_1}\Big)
\label{A-SCKM}
\end{eqnarray}  
From this equation we can get the $L$--$R$ down squark mass matrix
${m_{LR}^{2}}^{(d)}=v_1\ {Y^A_{d}}^* - \mu e^{i\varphi_{\mu}}
\tan\beta\, M_{d}$.
And finally using unitarity of $K$ we obtain for the $L$--$R$ 
Mass Insertions,
\begin{eqnarray}
\label{Dlr}
(\delta_{LR}^{(d)})_{i j}= \Frac{m_i}{m^2_{\tilde{q}}}\Big(
\delta_{ij}\ (c_{A} A_{C^9_3}^*\ +\ c_{\tilde{g}}\ m_{\tilde{g}}^*  ) -
\nonumber \\
\delta_{ij}\mu e^{i\varphi_{\mu}} \tan\beta +
K_{i 2}\ K^*_{j 2}\ c_{A}\ ( A_{C^9_2}^* - A_{C^9_3}^* ) +\nonumber \\
K_{i 3}\ K^*_{j 3}\ c_{A}\ ( A_{C^9_1}^* - A_{C^9_3}^* ) \Big)
\end{eqnarray}
where $m^2_{\tilde{q}}$ is an average squark mass and $m_i$ the quark mass.
The same rotation must be applied to the $L$--$L$ and $R$--$R$ squark mass 
matrices,
\begin{eqnarray}
{M^{(d)}_{LL}}^2 (M_W)=   K\ .\ M_Q^2 (M_W)\ .\ K^\dagger \nonumber \\
{M^{(d)}_{RR}}^2 (M_W)=   K\ .\ M_D^2 (M_W)\ .\ K^\dagger 
\end{eqnarray}
However, the off--diagonal MI in these matrices are sufficiently small
in this case thanks to the universal and dominant contribution from gluino 
to the squark mass matrices in the RGE. 

At this point, with the explicit expressions for $(\delta_{LR}^{(d)})_{i j}$,
we can study the gluino mediated contributions to EDMs and 
$\varepsilon^\prime/\varepsilon$. In this non--universal scenario, it is
relatively easy to maintain the SUSY contributions to the EDM of the
electron and the neutron below the experimental bounds while having
large SUSY phases that contribute to $\varepsilon^\prime/\varepsilon$.
This is due to the fact the EDM are mainly controled by flavor--diagonal
MI, while gluino contributions to $\varepsilon^\prime/\varepsilon$ are
controled by $(\delta_{LR}^{(d)})_{1 2}$ and $(\delta_{LR}^{(d)})_{2 1}$.
Here, we can have a very small phase for $(\delta_{LR}^{(d)})_{1 1}$
and $(\delta_{LR}^{(u)})_{1 1}$ and phases ${\cal{O}}(1)$ for the 
off--diagonal elements without any fine--tuning \cite{KKV}. 
It is important to remember that the observable phase is always the relative 
phase between these mass insertions and the relevant gaugino mass involved.
In Eq.(\ref{Dlr}) we can see that the diagonal elements tend to align 
with the gluino phase, hence to have a small EDM, it is enough to have
the phases of the gauginos and the $\mu$ term approximately equal, 
$\alpha_S=\alpha_1= - \varphi_\mu$. However $\alpha_2$ and $\alpha_3$ can still
contribute to off--diagonal elements. 
\begin{figure}[htb]
\vspace{9pt}
\epsfxsize = 8cm
\epsffile{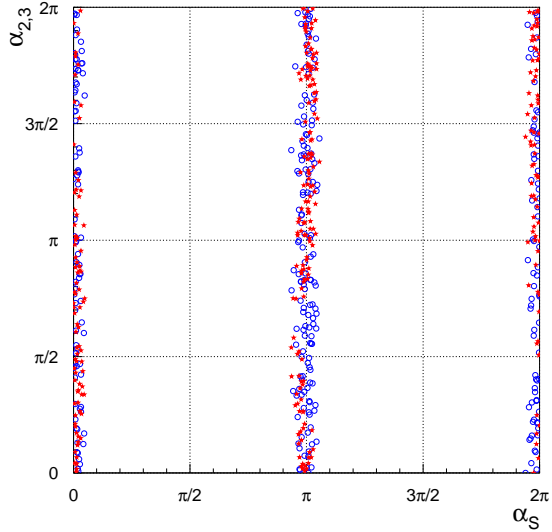}
\caption{Allowed values for $\alpha_2$--$\alpha_S$ (open blue circles) and 
$\alpha_3$--$\alpha_S$ (red stars)}
\label{scat}
\end{figure}
In figure \ref{scat} we show
the allowed values for $\alpha_S$, $\alpha_2$ and $\alpha_3$ assuming
$\alpha_1=\varphi_\mu=0$. We impose the EDM, $\varepsilon_K$ and 
$b\rightarrow s \gamma$ bounds separately for gluino and chargino 
contributions together with the usual bounds on SUSY masses.
We can see that, similarly to the CMSSM situation, $\varphi_\mu$ is 
constrained to be very close to the gluino and chargino phases
(in the plot $\alpha_S \simeq 0, \pi$), but $\alpha_2$ and 
$\alpha_3$ are completely unconstrained.
\begin{figure}[htb]
\vspace{9pt}
\epsfxsize = 8cm
\epsffile{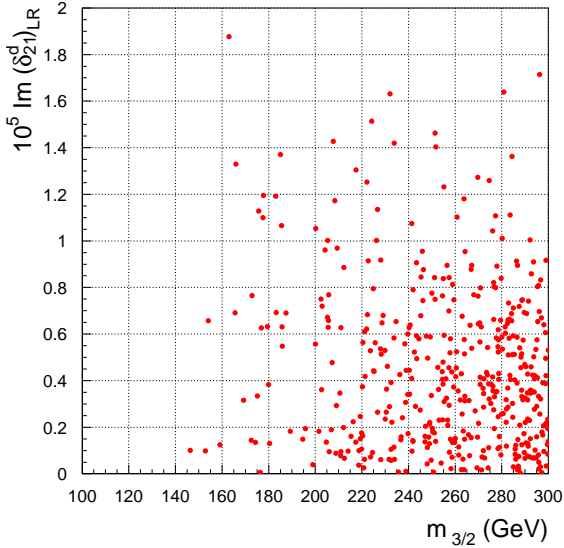}
\caption{$(\delta_{LR}^{(d)})_{2 1}$ versus $m_{3/2}$ for experimentally 
allowed regions of the SUSY parameter space}
\label{eps'}
\end{figure}
Finally, in figure \ref{eps'}, we show the effects of these phases in the
$(\delta_{LR}^{(d)})_{2 1}$ MI as a function of the gravitino mass.
All the points in this plot satisfy all CP--conserving constraints besides
EDM and $\varepsilon_K$ constraints. We must remember that a value of 
$|{\rm Im}(\delta^{d}_{12})_{LR}^{2}| \sim 10^{-5}$ gives a significant 
contribution to $\varepsilon^\prime/\varepsilon$. In this plot, we can see
a large percentage of points above or close to $1 \times 10^{-5}$. 
Hence, we can conclude that, in the presence of new flavor structures in 
the SUSY soft--breaking terms, it is not difficult to obtain sizeable SUSY 
contributions to CP violation observables and specially to
$\varepsilon^\prime/\varepsilon$ \cite{murayama,KKV}.\footnote{With these 
$L$--$R$ mass insertions alone, it is in general difficult to saturate
$\varepsilon_K$ \cite{gabbiani}. However, in some special situations, 
it is still possible to have large contributions \cite{brhlik2,isidori}.
On the other hand, the $L$--$L$ mass insertions can naturally contribute to
$\varepsilon_K$ \cite{MPV}}

\section{CONCLUSIONS AND OUTLOOK}

Here we summarize the main points of this talk:
\begin{itemize}
\item Flavor and CP problems constrain low--energy SUSY, but, at the same
time, provide new tools to search for SUSY indirectly.
\item In all generality, we expect new CP violating phases in the SUSY
sector. However, these new phases are not going to produce sizeable
effects as long as the SUSY model we consider does not exhibit a new flavor 
structure in addition to the SM Yukawa matrices.
\item In the presence of a new flavor structure in SUSY, we showed that 
large contributions to CP violating observables are indeed possible.
\end{itemize}

In summary, given the fact that LEP searches for SUSY particles are 
close to their conclusion and that for Tevatron it may be rather challenging
to find a SUSY evidence, we consider CP violation a potentially precious 
ground for SUSY searches before the advent of the ``SUSY machine'', LHC.

\end{document}